\documentclass[prb,twocolumn,english,superscriptaddress,footinbib]{revtex4-2}
\usepackage{amsmath,amssymb,mathrsfs}
\usepackage{amsfonts,bm}
\usepackage{amsmath}
\usepackage{amssymb}
\usepackage{amsthm}
\usepackage{graphicx}
\usepackage{graphics}
\usepackage{subfigure}
\usepackage{color}
\usepackage{bm}
\usepackage{hyperref}
\usepackage{rotating}
\usepackage{comment}
\usepackage[colorinlistoftodos]{todonotes}
\usepackage{esint}
\usepackage{xfrac}
\usepackage{natbib}
\usepackage{upgreek}
\usepackage{babel}
\usepackage{float}

\begin{document}

\title{Hydrodynamic Edge Modes and Fragile Surface States of Symmetry Protected Integer Quantum Hall Effect of Bosons}

\author{Dylan Reynolds}
\affiliation{Department of Physics, City College, City University of New York, New York, NY 10031, USA }
\affiliation{CUNY Graduate Center, New York, NY 10031}
\author{Gustavo M. Monteiro}
\affiliation{Department of Physics and Astronomy, College of Staten Island, CUNY, Staten Island, NY 10314, USA}
\author{Sriram Ganeshan}
\affiliation{Department of Physics, City College, City University of New York, New York, NY 10031, USA }
\affiliation{CUNY Graduate Center, New York, NY 10031}

\date{\today}

\begin{abstract}

We adapt the fluid description of Fractional Quantum Hall (FQH) states, as seen in Monteiro et al. (2022) ~\cite{monteiro2022topological}, to model a system of interacting two-component bosons. This system represents the simplest physical realization of an interacting bosonic Symmetry-Protected Topological (SPT) phase, also known as the integer quantum Hall effect (IQHE) of bosons. In particular, we demonstrate how the fluid dynamical boundary conditions of no-penetration and no-stress at a hard wall naturally give rise to the two counter-propagating boundary modes expected in these SPT phases. Moreover, we identify energy-conserving hydro boundary conditions that can either create a gap in these edge modes or completely isolate the edge states from the bulk, as described in Physical Review X 14, 011057 (2024), where they are termed fragile surface states. These fragile surface states are typically absent in K-matrix edge theories and require bulk dynamics to manifest. By leveraging insights from hydrodynamical boundary dynamics, we can further elucidate the intricate surface properties of SPTs beyond the usual topological quantum field theory based approaches.
\end{abstract}


\maketitle


\section{Introduction} \label{sec:introduction}

The discovery of topological insulators and superconductors has enlarged the notion of topological phases that owe their properties to symmetries ~\cite{hasan2010colloquium, qi2011topological, hasan2011three, konig2007quantum, hsieh2008topological, hsieh2009observation, hsieh2009first, xia2009observation, chen2009experimental, chen2011sciences, roushan2009topological, kane2005z, fu2007topological, bernevig2006quantum, fu2007topological, moore2007topological, roy2009topological, roy2006three, qi2008topological}. These topological phases are dubbed Symmetry-Protected Topological phases (SPTs), and their key features are bulk energy gaps and edge modes that are robust to symmetry-preserving perturbations. Subsequent works have generalized the ideas of SPT phases to quantum many-body states with interactions~\cite{ran2008spin, levin2009fractional, maciejko2010fractional, fidkowski2010effects, swingle2011correlated, neupert2011fractional, santos2011time, turner2011topological, gu2014symmetry, ryu2012interacting, ruegg2012topological, yao2013interaction, gu2014symmetry}. These interacting generalizations can be gapped short-range entangled without any intrinsic topological order but still possess robust edge modes and a bulk topological invariant. Interacting SPTs have been extensively classified using sophisticated mathematical tools such as group cohomology~\cite{gu2014symmetry} and topological quantum field theory methods such as K-matrix theory \cite{lu2012theory}.

Even though these frameworks capture all the essential topological features of the SPTs, it would be useful to quantify the microscopic dynamics in the context of Chern-Simons-Ginzburg-Landau (CSGL) field theories.  CSGL theory has been developed in the context of the fractional quantum Hall state by Zhang, Hansson, and Kivelson~\cite{zhang1989effective}, and independently by Read~\cite{read1989order}. The CSGL framework for SPTs is well understood, however the bosonic matter (GL) part of the CSGL theory is usually discarded while studying edge physics; typically the gauge invariance determines the boundary chiral dynamics in terms of additional edge fields. However, in the presence of bulk bosonic matter, the gauge invariance is preserved. Deriving the edge dynamics takes a different route, by utilizing the anomaly inflow principle, and does not need additional fields to be added to the edge. In our recent pieces of work~\cite{monteiro2022topological, monteiro2023coastal,monteiro2023kardar}, we have used this anomaly inflow mechanism to identify the superfluid boundary conditions that are consistent with the expected chiral edge dynamics. We also derived the non-linear generalization of this chiral boson action, where the chiral boson fields emerge from bulk fields taken at the boundary to satisfy the appropriate fluid dynamical boundary conditions. 

In this work, we generalize the anomaly inflow approach to derive a hydrodynamical model with appropriate boundary conditions of a particular interacting SPT phase, dubbed the Integer Quantum Hall Effect (IQHE) for bosons, introduced by Senthil and Levin in Ref.~\cite{senthil2013integer}. To arrive at an SPT phase, they start with a two-component system of bosons (spinor bosons or a bilayer system) in a large magnetic field and illustrate how this system has integer Hall conductivity if the $U(1)\times U(1)$ symmetry is preserved. The existence of two counter-propagating edge modes, one carrying charge and the other pseudospin, is derived using the K-matrix Chern-Simons formalism, albeit after dropping the bosonic matter in the bulk. Subsequent work has shown how these phases can manifest in interacting lattice models and two-component Bose gases \cite{he2015bosonic, furukawa2013integer}. Here we keep the bosonic matter and investigate how a hydrodynamic framework captures the bulk and edge properties of this SPT phase. We extract the bulk conductivity $\sigma_{xy}$ from the algebra of the fluid polarization of the total charge field, which is a uniquely hydrodynamical way of determining the bulk invariants.

In particular, we show how the fluid dynamical boundary conditions at the hard wall such as no-penetration and no-stress boundary conditions lead to counter-propagating chiral edge modes, one carrying charge and the other pseudospin. We also see the existence of two counter-propagating Kelvin Modes, non-dispersive modes that tend to accompany the chiral boson mode in fluid descriptions of quantum systems but are not associated with any anomaly. We then outline two types of energy-conserving boundary conditions that couple both edges at the boundaries without altering the bulk physics. The first type is the partial slip boundary condition, where the tangential stress in one layer generates slip in the second layer, and vice versa. These boundary conditions open a gap in the spectrum.

Remarkably, a second set of hydrodynamic boundary conditions results in the detachment of edge modes from the bulk. These isolated edge modes, which do not begin or end at the bulk bands, have recently been identified as fragile surface states in Ref.~\cite{altland2024fragility} for non-interacting topological insulators belonging to the non-Wigner-Dyson class. Within our framework, we demonstrate how the boundary conditions deform the edge 
$U(1)$ symmetry, leading to the decoupling of edge states from the bulk. We emphasize that fragile surface states are beyond the scope of traditional edge theories within topological quantum field theory, such as the K-matrix formalism, partly due to the absence of bulk matter. In contrast, within the hydrodynamic framework, the edge theory is consistently derived in conjunction with the bulk matter, which appears to be a requirement for uncovering the fragile surface states.

The benefits of a fluid dynamical approach to SPT phases are twofold. Firstly, it enables the systematic generalization of edge theories to include a richer class of surface phenomena, such as fragile states. Secondly, adopting a hydrodynamical approach may lead to the discovery of unique experimental signatures of the topological phase, imprinted in the non-universal matter dynamics accessible on ultracold atomic platforms. These platforms are likely where many of these phases will be realized in the near future.

\section{Mutual composite Boson theory} \label{sec:CSGL}

Following Ref. \cite{senthil2013integer}, we examine a two-dimensional system of two-component bosons (for example spinor bosons or a bilayer system) subject to a large magnetic field with short-ranged repulsive interactions. The large magnetic field ensures each component is in a $\nu=1$ integer quantum Hall phase. Ref.~\cite{senthil2013integer} considers a particular candidate state of this setup which is dubbed as an integer quantum Hall effect of bosons, which is a $U(1)$ symmetry-protected topological phase with Hall conductivity of $\sigma_{xy}=2 e^2/h$ and zero thermal Hall conductivity $\kappa_{xy}=0$. In the absence of tunneling between the two components, an additional $U(1)$ leads to a pseudo-spin Hall conductivity of $-2e^2/h$. 

Construction of such a candidate state was done using a two-component Chern-Simons-Ginzburg-Landau (CSGL) theory with a mutual Chern-Simons (CS) statistical term with a K-matrix of $K=\begin{pmatrix}
    0 & 1\\ 1 & 0
\end{pmatrix}$ implementing the flux
attachment. The role of the mutual CS term is to attach a flux quantum from one species to each boson of another species, resulting in a ``mutual composite boson fluid"  that experiences zero average flux.  Following the quantum Hall logic, this mutual-Chern-Simons-Ginzburg-Landau (mCSGL) effective action can be written as,
\begin{align}
S_{\text{bulk}} &=\int d^2x\, dt\left[\sum_{a}\mathcal{L}_{a} + \mathcal{L}_{\text{int}} + \mathcal{L}_{\text{CS}} \right] , \label{eq:csglaction}
\end{align}
where the two individual component Lagrangians associated with the bosonic matter of each species, labeled by $a=1,2$, are given by
\begin{align}
\mathcal{L}_{a} &=i\hbar\left(\Phi^a\right)^\dagger D_t^{a}\Phi^a - \frac{\hbar^2}{2m} \left|D_i^{a}\Phi^a\right|^2 ,
\end{align}
the mutual Chern Simons (CS) Lagrangian is given by
\begin{align}
\mathcal{L}_{\text{CS}} &= \frac{\hbar}{4\pi}\sum_{a,b=1}^2\epsilon^{\mu\nu\lambda}K^{ab}\alpha^{a}_\mu \partial_\nu \alpha^{b}_\lambda . \label{eq:CSlagrangian}
\end{align}
and the interaction Lagrangian $\mathcal L_{\text{int}}$ is solely a function of $|\Phi^{a}|^2$. A minimal coupling to the external electromagnetic vector potential $A_\mu$ and internal Chern-Simons statistical fields $\alpha^{a}_\mu$ are included in the covariant derivatives, defined as $ D^{a}_\mu=\partial_\mu - i\frac{q}{\hbar}A_\mu + i \alpha^{a}_\mu $. Both species have the same effective mass $m$ and charge $q$. Greek indices $\mu, \nu, \lambda$ run over $t,x,y$, while Latin indices $i,j,k$ run over the spatial components $x,y$. 

The interaction term is assumed to introduce a non-zero vacuum expectation value for both species. This can be approximated by a local repulsive interaction, due to density fluctuation on top of a uniform density background, i.e., the jellium model, with local interactions. Therefore, we can express the interaction Lagrangian as
\begin{align}
\mathcal{L}_{\text{int}} &= -\sum_{a,b} V_{ab}\left(|\Phi^a|^2-\frac{qB}{2\pi\hbar}\right)\left(|\Phi^b|^2-\frac{qB}{2\pi\hbar}\right).
\end{align}
Here, we are assuming that both fields have the same vacuum expectation value, which is given by $qB/(2\pi\hbar)$, where $B$ is the external magnetic field. This leads to two copies of an abelian Higgs mechanism, which can be seen explicitly if we express the scalar fields in their polar forms, that is,
\begin{align}
   \Phi_{a}&=\sqrt{\frac{qB}{2\pi\hbar}}\left(1+\frac{n^{a}}{2}\right)e^{i\theta^{a}} \,. \label{eq:Madelung}
\end{align}
The factor of $\tfrac{1}{2}$ was introduced to give us $|\Phi^{a}|^2-\tfrac{qB}{2\pi\hbar}\approx \tfrac{qB}{2\pi\hbar}n^{a}$. Using the Madelung variables defined in Eq.~(\ref{eq:Madelung}), we see that the Lagrangian, up to quadratic order, becomes
\begin{align}
    \mathcal L^{(2)}_{a}=&-\frac{qB}{2\pi}\left[n^{a}\left(\partial_t\theta^{a}+\alpha_0^{a}-\frac{q}{\hbar}A_0\right)+\frac{\hbar}{8m}(\partial_i n^{a})^2\right.\nonumber
    \\
    &+\left.\frac{\hbar}{2m}\left(\partial_i\theta^{a}+\alpha_i^{a}-\frac{q}{\hbar}A_i\right)^2+\alpha_0^a\right],
    \\
    \mathcal L^{(2)}_{\text{int}}= &-\frac{q^2B^2}{4\pi^2\hbar^2}\sum_{a,b} V^{ab} \,n^a n^b\,.
\end{align}
Note that $\mathcal L_{\text{CS}}$ is already quadratic in fields.

The linearized equations of motion are thus
\begin{align}
\partial_t\theta^{a}+\alpha_0^{a}-\frac{q}{\hbar}A_0-\frac{\hbar}{4m}\nabla^2 n^{a}+\frac{qB}{\pi\hbar^2}\sum_b V^{ab} n^{b} &= 0\,, \label{eq:theta}
\\
\partial_t n^{a}+\frac{\hbar}{m}\partial_i\left(\partial_i\theta^{a}+\alpha_i^{a}-\frac{q}{\hbar}A_i\right) &= 0\,, \label{eq:n}
\\
\frac{qB}{2\pi}\left(1+n^{a}\right) - \frac{\hbar}{2\pi}\epsilon_{ij}\sum_{b}K^{ab}\partial_i\alpha_j^{b} &= 0\,, \label{eq:Gauss}
\\
\frac{qB}{ m}\left(\partial_i\theta^{a}+\alpha_i^{a}-\frac{q}{\hbar}A_i\right) +\epsilon_{ij}\sum_{b}K^{ab}\left(\partial_t\alpha_j^{b}-\partial_j\alpha_0^b\right) &= 0\,. \label{eq:alpha}
\end{align}
Here we have introduced $\epsilon_{ij}$ as the antisymmetric tensor in 2D and employed the K-matrix $K^{ab}$. Note that the above system constitutes two sets of equations, one for each species. For simplicity, we have assumed a uniform $B$ field.

From here on, our analysis differs from that presented in \cite{senthil2013integer}. We focus on the superfluid hydrodynamics of bosonic matter subject to vorticity constraints enforced by the Chern-Simons terms in the presence of boundaries. This approach deviates from the traditional strategy, which involves considering the effective Chern-Simons theory without any bulk matter and deducing boundary dynamics through the enforcement of gauge invariance, which requires additional gapless degrees of freedom. The key result of this paper is that we derive the bulk and boundary topological properties directly from the superfluid hydrodynamics. This follows our recent work, which employed a similar strategy for a Laughlin state described by a CSGL action~\cite{monteiro2022topological, monteiro2023coastal, monteiro2023kardar} and is in the same spirit as M. Stone's hydrodynamic interpretation of CSGL saddle point equations~\cite{stone1990superfluid}.

\section{Bulk topological invariants from the algebra of fluid polarization}

The governing equations of this system (\ref{eq:theta})-(\ref{eq:alpha}) admit an alternative formulation in terms of the fluid polarization.  To construct the fluid polarization, we recognize $\tfrac{q^2B}{2\pi\hbar}n^{a}$ as plasmon fluctuations of the model. These fluctuations then can be expressed in terms of polarization waves, under the identification 
\begin{align}
    \frac{q^2B}{2\pi\hbar}n^{a}=-\partial_iP_i^{a}\,,
\end{align}
where $P_i^{a}$ is the polarization field. Using equation~(\ref{eq:Gauss}) and imposing that the polarization field must be gauge invariant, we find that
\begin{align}
P_i^a = \frac{q}{2\pi}\epsilon_{ij}\sum_{b}K^{ab}\left(\partial_j\theta^b +\alpha^b_j - \frac{q}{\hbar}A_j \right) . \label{eq:polarizationdef}
\end{align}
That is, the polarization of one species is defined solely in terms of the opposite species. This can be made precise by shifting the Chern-Simons gauge field, that is, $\alpha_\mu^{a}\rightarrow \alpha_\mu^{a} +\tfrac{q}{\hbar}A_\mu$ and identify the terms of the form $P_i^a E_i$, where $E_i$ is the external electric field. For more details, we refer to our previous work~\cite{monteiro2022topological}.  

We can read the polarization algebra directly from the symplectic structure of the mutual Chern-Simon action, which gives us
\begin{align}
\left\{P_i^{a}(\vec{x})\,,P_i^{b}(\vec{x}\,')\right\} = \frac{q^2}{2\pi\hbar}\epsilon_{ij}K^{ab}\,\,\delta(\vec{x}-\vec{x}\,')\,. \label{eq:P-algebra}
\end{align}
The polarization fields require the matter term $\partial_i\theta^{a}$ to ensure the consistency of the algebra between the polarization and the density fields, that is,
\begin{align}
    \{n^a(\vec x),P_i^b(\vec x\,')\}=-\frac{2\pi\hbar}{q^2B}\{\partial_iP_i^{a}(\vec x),P_j^{b}(\vec x\,')\}.
\end{align}

We can decouple the polarization algebra by diagonalizing the K-matrix, which naturally introduces the polarization vectors for charge and pseudospin
\begin{align}
P_i^{Q} = P_i^1 + P_i^2 ,\quad P_i^{S} = P_i^1 - P_i^2  .
\end{align}
As we will see in section \ref{sec:chargespin}, the same logic applied to the matter fields decouples the bulk equations. The decoupled polarization algebra is then expressed as 
\begin{align}
\left\{P_i^{(Q)}(\vec{x})\,,P_j^{(Q)}(\vec{x}\,')\right\} &=  2\frac{q^2}{2\pi\hbar}\epsilon_{ij}\delta(\vec{x}-\vec{x}\,'), \\
\left\{P_i^{(S)}(\vec{x})\,,P_j^{(S)}(\vec{x}\,')\right\} &=  -2\frac{q^2}{2\pi\hbar}\epsilon_{ij}\delta(\vec{x}-\vec{x}\,'), \\
\left\{P_i^{(Q)}(\vec{x})\,,P_j^{(S)}(\vec{x}\,')\right\} &= 0 .
\end{align}

Since polarization is crucially linked to the geometric Berry phase and associated Hall conductivity \cite{resta1994macroscopic, ortiz1994macroscopic, resta2007theory, resta2010electrical}, we can immediately read off the magnitude of the charge and pseudospin Hall conductivities, $\sigma_{xy}^{(Q)}=-\sigma_{xy}^{(S)}=2\frac{q^2}{2\pi\hbar}$ thereby quantifying the bulk invariant from in terms of fluid variables. In the subsequent sections, we extract the edge dynamics within the hydrodynamic equation by identifying the superfluid boundary conditions at the edge that are consistent with the anomalous edge dynamics and the bulk topological properties.

\section{First order hydrodynamics and the choice of velocity field} \label{sec:linear}

The equations defined in Eqs.~(\ref{eq:theta})-(\ref{eq:alpha}) can be written in the form of hydrodynamic equations by identifying the velocity field. For the Laughlin state, the superfluid formulation was introduced by Stone in Ref~\cite{stone1990superfluid} leading to continuity and Euler equations. Within Stone's FQH fluid dynamics, the Euler equations possessed three derivatives of density fields (two derivatives of density in the stress tensor), which are the so called ``quantum pressure" terms. However, our recent works ~\cite{abanov2013fqhefluid, monteiro2022topological} have shown that one can change the velocity field definition, such that, the Euler equation posses only second-order derivatives (one derivative of velocity in the stress tensor). This choice does not alter the bulk properties. The upshot is that the two physical boundary conditions applied to this velocity field will assume the familiar fluid dynamical forms of no-penetration, combined with either no-stress or no-slip.

In the case of the bosonic Integer Quantum Hall (IQH), the velocity fields for the two components can be defined as follows:
\begin{align}
 v_i^{a}=\frac{\hbar}{m}\left(\partial_i\theta^{a}+\alpha_i^{a}-\frac{q}{\hbar}A_i-\frac{1}{2}\sum_b K^{ab}\epsilon_{ij}\partial_j n^{b}\right) \label{eq:vdef}    
 \end{align}
In terms of this velocity field, the system (\ref{eq:theta})-(\ref{eq:alpha}) becomes
\begin{align}
\epsilon_{ij}\partial_i v_j^a - \frac{1}{2}\omega_B\ell_B^2\nabla^2 n_a -  \omega_B \sum_b K^{ab} n^b  &=0\,,\label{eq:LinearHall}\\
\partial_t n^a + \partial_i v_i^a &= 0\,, \label{eq:LinearCont} \\
\partial_t v_i^a - \partial_j T_{ij}^a  -  \omega_B\epsilon_{ij}\sum_b K^{ab}v_j^b&=0\,\label{eq:LinearMomentum},
\end{align}
where we've introduced the length and time scales set by the magnetic length $\ell_B^2=\tfrac{\hbar}{qB}$ and cyclotron frequency $\omega_B=\tfrac{qB}{m}$. The linearized stress tensor is given by
\begin{align}
T_{ij}^a = -\delta_{ij}P^a + \frac{1}{2}\omega_B\ell_B^2 \sum_b K^{ab}\left(\epsilon_{ik}\partial_k v_j^b + \epsilon_{jk}\partial_i v_k^b \right) , \label{eq:linearstress}
\end{align}
with pressure
\begin{align}
P^a =\ell_B^2\omega_B^2\, n^a + \frac{1}{\pi m \ell_B^2}\sum_{b} V^{ab} n^b ,
\end{align}
where we have used that $V^{ab}=V^{ba}$, which follows directly from $\mathcal L_{\text{int}}$.

Note that the Lorentz force term in one system is sourced by the velocity of the other species. Additionally, the form of the stress tensor (\ref{eq:linearstress}) suggests that the off-diagonal hydrodynamic stresses exerted on one species originate entirely from the flow of the opposing species. Furthermore, this stress tensor takes the form of classical odd viscosity ~\cite{monteiro2021hamiltonian, monteiro2022topological}, which stems from our definition of velocity~\footnote{This is also true for all two-dimensional superfluids modeled by a Gross-Pitaevskii equation in terms of the Madelung variables, where the quantum pressure terms written in the standard velocity definition of $u_i=\partial_i\theta$ takes the form of odd viscosity in terms of $v_i=(\hbar/m)(\partial_i \theta-\frac{1}{2}\epsilon_{ij}\partial_j n_j)$ }.

\section{Chiral edge modes from fluid dynamical boundary conditions} \label{sec:boundaryconditions}

In addition to bulk equations, a fluid system must be accompanied by appropriate boundary conditions. In principle, there exists a family of boundary conditions that correspond to different physical scenarios, but typically we choose ones that capture the observed or expected physics near the boundary. This is also the case in conventional fluid dynamics where we pick no-penetration and no-slip boundary conditions (zero velocity) when we study the motion of a solid body in water and no-stress (force balance) at two-fluid interfaces such as oil in water. Note that two boundary conditions are required to consistently solve for the fields with second-order derivatives in the equations of motion. For quantum fluids defined in Eqs.~(\ref{eq:LinearHall}-\ref{eq:linearstress}) in their ground state, we enforce that the boundary conditions are energy-conserving. Even though energy dissipation can be introduced at the boundaries in some restricted sense~\cite{monteiro2023kardar}, we will not consider boundary conditions that do not conserve energy. 

To this end, we consider the additional energy conservation equation
\begin{align}
\partial_t \mathcal{H} + \partial_i \mathcal{Q}_i =0 , \label{eq:energycontinuity}
\end{align}
where $\mathcal{H}$ contains typical kinetic terms, as well as any additional potential energy terms and $\mathcal{Q}_i$ is the energy current. To analyze the boundary conditions we take the fluid domain to be the lower half plane $y \leq 0$, with a rigid interface along the $x$ axis. Once $\mathcal{H}$ and $\mathcal{Q}_i$ are identified, we can enforce conservation of energy including both bulk and boundary terms in the following way
\begin{align}
\frac{dE}{dt} = \int d^2 x\, \partial_t \mathcal{H} = -\int d^2\,\partial_i\mathcal{Q}_i = -\int dx\,\mathcal{Q}_y \Big|_{y=0} \,,
\end{align}
where we've used the divergence theorem and assumed all quantities vanish far from the boundary. For energy to be conserved we enforce that
\begin{align}
\mathcal{Q}_y \Big|_{y=0} = 0 . \label{eq:generalbc}
\end{align}
Along with particle number (or equivalently, mass or charge) conservation, this gives the second boundary condition required in a second-order system.

The Hamiltonian for our linear system is~\footnote{Note our Hamiltonian does not possess any terms of the form $\sim (\partial_i n)^2$, which typically arise in a linear superfluid Hamiltonian. In the two-fluid case, these terms cancel, as can be checked by examining the full nonlinear theory.}
\begin{align}
\mathcal{H} &= \sum_{a}\left(\frac{m}{2} v_i^av_i^a  + \frac{m}{2}\omega_B^2\ell_B^2 \left(n^a\right)^2 \right) +\frac{1}{2\pi \ell_B^2} \sum_{ab}V^{ab}n^a n^b . \label{eq:linearVham}
\end{align}
Using the equations of motion we find the corresponding conserved current, satisfying (\ref{eq:energycontinuity}), to be
\begin{align}
\mathcal{Q}_j =  - m \sum_{a}v_i^a T_{ij}^a .
\end{align}
The no-energy dissipation condition in Eq. (\ref{eq:generalbc}) imposes the following constraint,
\begin{align}
\mathcal{Q}_y\Big|_{y=0} =  - m \sum_{a} (v_x^aT_{xy}^a+ v_y^aT_{yy}^a) =0. \label{eq:qyspec}
\end{align}
Within the above restriction, we can deduce energy-preserving boundary fluid dynamical boundary conditions.

One standard choice is the no-penetration condition, which says that the fluid does not flow into a hard wall
\begin{align}
  v_y^a\Big|_{y=0} =0  .
\end{align}
This eliminates the second term of (\ref{eq:qyspec}). The second boundary condition must then force the first term to vanish. In fact, there are two separate classes of boundary conditions that allow this to happen. The first case (Case I) is known as the partial slip condition, and can be expressed as
\begin{align}
 T_{xy}^{1}\Big|_{y=0}= - \lambda v_x^{2}\Big|_{y=0}\,,\quad T_{xy}^{2}\Big|_{y=0}=\lambda v_x^{1}\Big|_{y=0} \,. \label{eq:lambdacond}
\end{align}
The parameter $\lambda$ corresponds to an inverse slip length and it interpolates between the no-stress condition for $\lambda=0$ to the no-slip condition for $\lambda\rightarrow\infty$. For the intermediate values of $\lambda$ the energy conservation requires that the tangent stress of one component generates a partial slip in the second component. The second class of boundary conditions (Case II) is given by,
\begin{align}
    v_x^{1}\Big|_{y=0}=-\gamma v_{x}^2\Big|_{y=0}\,,\quad T_{xy}^2\Big|_{y=0}=\gamma T_{xy}^1\Big|_{y=0}\,. \label{eq:gammacond}
\end{align}
This condition, parameterized by $\gamma$, matches the tangent velocity and the tangent stress of the two layers.

From the standpoint of energy conservation, all these boundary conditions are equally valid, though they result in different edge physics. We now investigate how these boundary conditions encapsulate the anomaly inflow mechanism of the topological phase. In our recent work, we demonstrated that if the system is anticipated to exhibit anomaly-induced chiral modes that propagate along the boundary, the no-stress condition is preferred over the no-slip condition. This is because the no-slip condition by definition forbids any chiral dynamics along the edge whereas the no-stress condition results in chiral edge dynamics induced by the anomaly inflow mechanism \cite{monteiro2022topological}.

We first consider the two-component generalization of the boundary conditions considered for the Laughlin state in Ref.~\cite{monteiro2022topological}. These conditions correspond to the $\lambda=0$ limit of the partial slip conditions and are given by,
\begin{align}
 v_y^{a}\Big|_{y=0} =0\,, \quad    T_{xy}^{a}\Big|_{y=0}=0\,. \label{eq:bc}
\end{align}
Using the continuity Eq.~(\ref{eq:LinearCont}) combined with the no-penetration condition, we observe that the no-stress condition can be written in a dynamical form as
\begin{align}
 \Big[\partial_t n^a + 2 \partial_x v_x^a\Big]_{y=0}=0 \,. \label{eq:tyx}
\end{align}
In Sec.~\ref{sec:mode}, we solve for the bulk and edge dispersion for the hydro equations (\ref{eq:LinearHall}-\ref{eq:linearstress}) together with boundary conditions (\ref{eq:bc}). We then consider the most general boundary conditions and show how the edge dynamics change as a function of the boundary parameters $\lambda$ and $\gamma$. In Sec.~\ref{sec:BC-action} we construct an effective action for the both cases, which requires the addition of an auxiliary chiral boson field at the edge to obtain the correct boundary conditions. 

\section{Charge and Pseudospin Basis} \label{sec:chargespin}

The linear system (\ref{eq:LinearHall}-\ref{eq:LinearMomentum}) is naturally coupled, due to the structure of the K-matrix. We can decouple them into two independent systems by introducing the charge and pseudospin, with densities and velocities defined as
\begin{align}
\rho^Q &= n^1 + n^2 , \quad \rho^S = n^1 - n^2 , \label{eq:Sdef}\\
V_i^Q &= v_i^1 + v_i^2 , \quad V_i^S = v_i^1 - v_i^1 . \label{eq:Udef}
\end{align}
Additionally, we take each species to have the same self-interaction energy $V^{11}=V^{22}$. This assumption can be relaxed but requires modification to the above definitions. The above linear transformation will decouple the system, and give the resulting modes a physical meaning; modes that carry charge, and modes that carry pseudospin. In these variables, we have two decoupled subsystems 
\begin{align}
\epsilon_{ij}\partial_i V_j^\alpha + \hat{L}^\alpha\rho^\alpha &= 0 \,, \label{eq:hallQ}\\
\partial_t \rho^\alpha + \partial_i V_i^\alpha &= 0\,,\label{eq:Qcont}\\
\partial_t V_i^\alpha + (c^\alpha)^2\partial_i \rho^\alpha + \hat{L}^\alpha\epsilon_{ij}V_j^\alpha &= 0 , \label{eq:Qmomentum}
\end{align}
where $\alpha=Q,S$. For brevity we've introduced the operator $\hat{L}^Q =-\omega_B \left(1+\tfrac{1}{2}\ell_B^2\nabla^2\right)$ for the charge system, and $\hat{L}^S =\omega_B \left(1+\tfrac{1}{2}\ell_B^2\nabla^2\right)$ for the pseudospin system. The only difference between the two systems is the sign of $\omega_B$. We've also defined
\begin{align}
(c^Q)^2 &= \frac{1}{\pi m \ell_B^2}\left(V^{11} + V^{12}\right) +  \ell_B^2\omega_B^2,\\
(c^S)^2 &= \frac{1}{\pi m \ell_B^2}\left(V^{11} - V^{12}\right) +  \ell_B^2\omega_B^2 ,
\end{align}
as the sound velocity associated with the charge and pseudospin, respectively (recall that with our assumptions $V^{11}=V^{22}$ and $V^{12}=V^{21}$).

The boundary conditions can also be recast in the charge and pseudospin variables. For $\lambda=0$ in Case I, the boundary conditions remained decoupled
\begin{align}
V_y^\alpha \Big|_{y=0} &= 0 , \label{eq:Vnopen} \\
\Big[ \partial_t \rho^\alpha + 2 \partial_x V_x^\alpha\Big]_{y=0} &=0 . \label{eq:Vnostress} 
\end{align}
For a general $\lambda$ value we still retain the no-penetration condition (\ref{eq:Vnopen}), but the form of (\ref{eq:lambdacond}) implies the no-stress condition is replaced by
\begin{align}
\frac{1}{2}\omega_B\ell_B^2\Big[ \partial_t \rho^\alpha + 2 \partial_x V_x^\alpha\Big]_{y=0} &=-\lambda \sum_\beta \epsilon^{\alpha\beta} V_x^\beta\Big|_{y=0}. \label{eq:lambdabc}
\end{align}
Physically this boundary condition implies that the stress generated by the edge charge (spin) generates a slip length for the spin (charge). It's also clear that nonzero $\lambda$ spoils the edge continuity equation, and as we will see, gaps out the edge modes. For Case II, we likewise keep the no-penetration condition,  but the charge/spin variables (\ref{eq:gammacond}) can be rewritten as
\begin{align}
(\gamma+1)V_x^Q=(\gamma-1)V_x^S , \label{eq:gammaV}
\end{align}
along with a continuity equation of the form
\begin{align}
\partial_t \Big[(1-\gamma)\rho^Q &+(1+\gamma)\rho^S\Big] \nonumber \\
& + 2\partial_x \Big[(1-\gamma)V_x^Q+(1+\gamma)V_x^S\Big]=0 . \label{eq:gammacont}
\end{align}
This indicates that one of the emergent edge $U(1)$ symmetry is maintained for this class of boundary conditions, and a gapless edge mode is still present. However, an interesting aspect of this $U(1)$ symmetry is that the $\gamma$ coefficient can deform the edge charge and current in a way that it need not respect spectral flow conditions with either the charge or pseudospin bulk bands. Consequently, this case leads to the so-called fragile surface states~\cite{altland2024fragility} that live as a separate band and do not begin or end at the bulk bands except at $\gamma=\pm 1$.

\section{Mode Structure} \label{sec:mode}

In this section, we explicitly solve for the mode structure of bulk and boundary, derived from the fluid dynamical boundary conditions. We find the bulk modes of the system by expanding the fields $\vec{\eta}\,^\alpha=(n^\alpha,V_x^\alpha,V_y^\alpha)$ as
\begin{align}
\vec{\eta}\,^\alpha = \int d\omega\, d^2 q\,\, \tilde{\vec{\eta}}\,^\alpha e^{-i\omega t + i \vec{q}\cdot\vec{x}} ,
\end{align}
which readily gives the bulk dispersion
\begin{align}
\omega^\alpha &= \pm\sqrt{(c^\alpha)^2 q^2 + \omega_B^2\left(1 - \frac{1}{2}\ell_B^2 q^2\right)^2} \label{eq:bulkdispcomp},
\end{align}
and corresponding eigenvectors
\begin{align}
\tilde{\vec{\eta}}\,^\alpha  = (q^2 \,,\, \omega q_x - iq_y L^\alpha \,,\, \omega q_y + iq_x L^\alpha )\,, \label{eq:eigenvec}
\end{align}
where $L^Q = -\omega_B\left(1 - \frac{1}{2}\ell_B^2 q^2\right)$ for the $Q$ system, and $L^S = \omega_B\left(1 - \frac{1}{2}\ell_B^2 q^2\right)$ for the $S$ system. The structure of these bulk bands is the same for each subsystem, the only difference being the sound velocity. To prevent interband mixing and guarantee our state remains in the lowest Landau level, the original potentials must satisfy
\begin{align}
V^{11} + V^{12} & < 2\pi\hbar\ell_B\omega_B, \quad V^{11} - V^{12}  < 2\pi\hbar\ell_B\omega_B.
\end{align}
Additionally, we must have $V^{11} + V^{12}  \geq 0$ and $V^{11} - V^{12}  \geq 0$ for this state to be a stable minimum (again recall that with our assumptions $V^{11}=V^{22}$ and $V^{12}=V^{21}$).

We now show how the expected edge modes arise from the hydrodynamic boundary conditions.  For details of the edge mode calculations, we refer the reader to \cite{monteiro2023coastal}. First, we expand the fields in modes localized near the hard wall boundary at $y=0$,
\begin{align}
\vec{\eta}\,^\alpha 
= \int d\omega d k\, \sum_{\sigma=1}^{2} C_\sigma^\alpha\, \tilde{\vec{\eta}}\,^\alpha  e^{-i\omega t + i k x + s_\sigma y}, \label{eq:mode1}
\end{align}
where $C_\sigma^\alpha$ is some expansion coefficient. Here $s_\sigma$ is a solution to the polynomial equation that arises by taking $(q_x,q_y) \rightarrow (k,-is)$ in (\ref{eq:bulkdispcomp}). Importantly, we require that $s$ have a positive real part to guarantee solution decay into the bulk (lower half-plane). The quartic polynomial admits exactly two roots with positive real parts, leading to the two terms in (\ref{eq:mode1}). The eigenvectors take the same form as (\ref{eq:eigenvec}), with the same replacement $(q_x,q_y) \rightarrow (k,-is)$. 

First, we apply the no-penetration boundary conditions (\ref{eq:Vnopen}). It's straightforward to show that this can be satisfied by setting $V_y^\alpha=0$ in the entire fluid domain and taking the edge dispersions to be
\begin{align}
\omega^Q =-c^Q k ,\quad \omega^S =c^S k .
\end{align}
These are known as Kelvin modes, in analogy with the coastal Kelvin modes present at the boundary of the shallow water model of ocean waves \cite{delplace2017topological, tauber2019bulk, tong2023gauge, monteiro2023coastal}. These modes are naturally non-dispersive, and as expected for an SPT phase, they are counter-propagating. A charge Kelvin wave propagates in the negative $x$ direction, while a pseudospin Kelvin wave propagates in the positive $x$ direction. The direction of these waves is ultimately set by the original external magnetic field, but is protected by the $U(1) \times U(1)$ symmetry associated with charge and pseudospin conservation.

Alternatively, for the appropriate choice of expansion coefficients $C_\sigma^\alpha$, we can satisfy the no-penetration boundary condition while keeping a non-vanishing $V_y^\alpha$ in the bulk. This leaves only a single overall amplitude, that allows us to write $\tilde{V_x}^\alpha$ at the boundary in terms of $\tilde{n}^\alpha$. Comparing the components of the eigenvectors we can write
\begin{align}
\tilde{V_x}^\alpha &= \chi^\alpha(\omega, k)\, \tilde{n}^\alpha , \label{eq:phidef}
\end{align}
where $\chi^\alpha(\omega, k)$ is a ratio of $\tilde{V}_x^\alpha$ and $\tilde{n}^\alpha$ evaluated at $y=0$. Again, for explicit details of this calculation see \cite{monteiro2023coastal}. To gain insight into the behavior of the modes we can expand $\chi^\alpha(\omega, k)$ for small $\omega$ and $k$
\begin{align}
\chi^Q(\omega, k) &= -c^Q \left(1 - \frac{1}{4}\ell_B^2 k^2 + \cdots \right) \label{eq:Qphi},\\
\chi^S(\omega, k) &= c^S \left(1 - \frac{1}{4}\ell_B^2 k^2  + \cdots \right). \label{eq:Sphi}
\end{align}
The above relations are useful for analyzing the various second boundary conditions (\ref{eq:Vnostress}-\ref{eq:gammacont}). We now split the analysis into cases.

\subsection{Case I with $\lambda=0$ (No-Stress)}

\begin{figure}[H]
\centering
\includegraphics[scale=0.46]{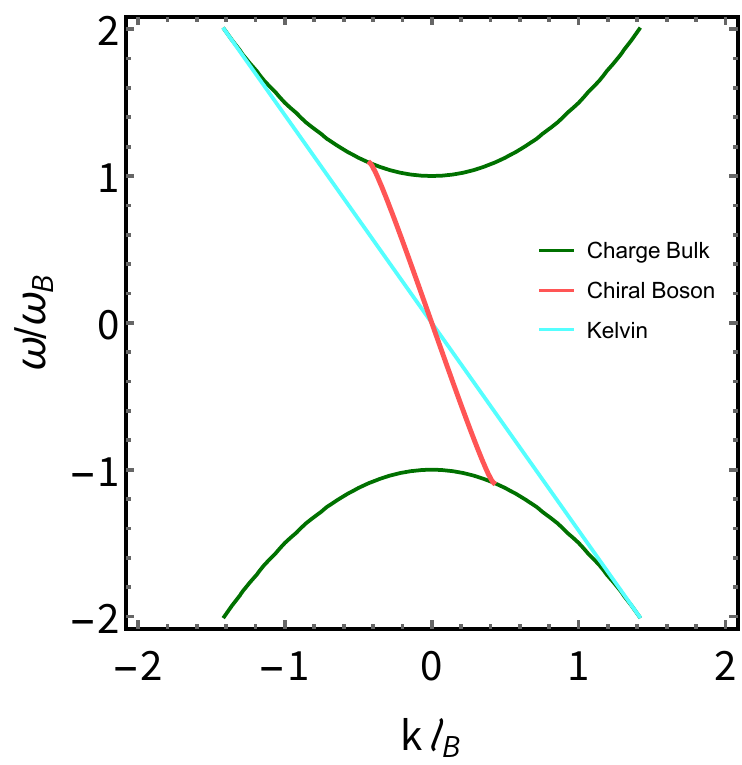}
\includegraphics[scale=0.46]{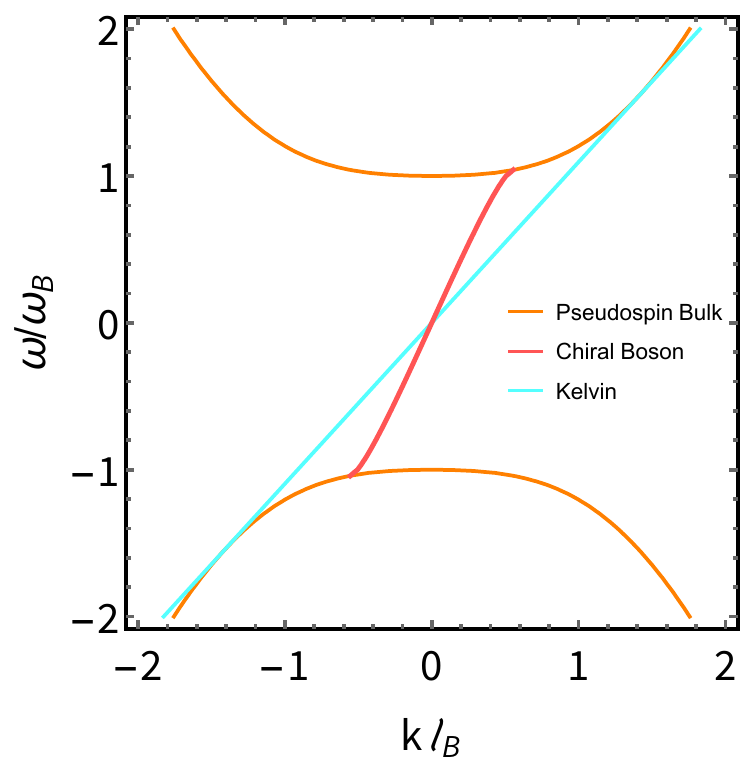}
\caption{Charge and pseudospin bulk bands, Kelvin modes and $\lambda=0$ chiral bosons, with $V^{11}=V^{22}=0.6\pi\hbar\omega_B\ell_B$, and $V^{12}=V^{21}=0.4\pi\hbar\omega_B\ell_B$.}
\label{fig:Plot}
\end{figure}

First, we look at the no-stress condition which gives us the expected edge modes predicted in \cite{senthil2013integer}. Using the relations (\ref{eq:phidef}), (\ref{eq:Qphi}), and (\ref{eq:Sphi}), we find the following expansion for the edge mode dispersions
\begin{align}
\omega^Q &= -c^Q\left(2 k - \frac{\ell_B^2}{2} k^3 +\cdots\right) ,\label{eq:Qboson}\\
\omega^S &= c^S\left(2 k - \frac{\ell_B^2}{2} k^3 +\cdots\right) \label{eq:Sboson} .
\end{align}
Following the terminology of the single fluid FQH analysis \cite{monteiro2023coastal} we call these chiral boson modes. As with the Kelvin modes they counter-propagate, with the charged mode moving in the negative $x$ direction, and the pseudospin mode moving in the positive $x$ direction. For a full numerical solution of the Kelvin mode and the $\lambda=0$ chiral boson see Fig. \ref{fig:Plot}.

\subsection{Case I with general $\lambda$: Gapped surface states}

\begin{figure}[H]
\centering
\includegraphics[scale=0.415]{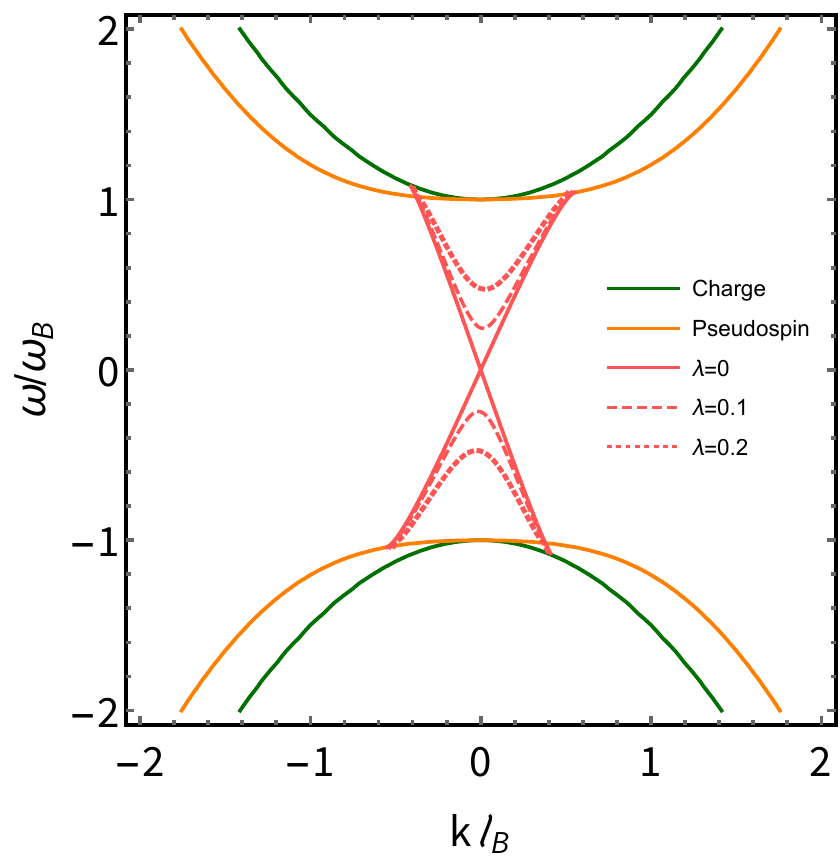}
\caption{Edge modes for various values of $\lambda$, with $V^{11}=V^{22}=0.6\pi\hbar\omega_B\ell_B$, and $V^{12}=V^{21}=0.4\pi\hbar\omega_B\ell_B$. Note that the two Kelvin modes are also present, but we omit it from the plot.}
\label{fig:Lambda}
\end{figure}

Next, we analyze the more general condition for $\lambda \neq 0$. Using (\ref{eq:phidef}) we write (\ref{eq:lambdabc}) as a $2$x$2$ system in Fourier space for the variables $\tilde{\rho}^Q$ and $\tilde{\rho}^S$. The determinant of this system gives the defining relation
\begin{align}
\left(\omega - 2 k \chi^Q\right)\left(\omega - 2 k \chi^S\right)+\frac{4\lambda^2}{\omega_B^2\ell_B^4} \chi^Q\chi^S = 0 .
\end{align}
For $\lambda=0$ we clearly recover (\ref{eq:Qboson}) and (\ref{eq:Sboson}). Remarkably, for nonzero $\lambda$ we find that these modes are gapped, with a bandgap of size $\omega_0=\tfrac{4 \lambda}{\omega_B\ell_B^2}\sqrt{c^Q c^S}$. In Fig. \ref{fig:Lambda} we give the full numerical solution, where we clearly see that increasing values of $\lambda$ increases the gap size, and the two chiral boson modes, discussed above, are joined. For small $\omega$ and $k$ we may use (\ref{eq:Qphi}), and (\ref{eq:Sphi}) which gives us the following leading order dispersion
\begin{align}
\omega = \pm \frac{2 \lambda}{\omega_B\ell_B^2}\sqrt{c^Q c^S} - \left(c^Q-c^S\right) k + \cdots .
\end{align}
Note that $c^Q-c^S$ is only nonzero in the presence of the off diagonal potential term $V^{12}$. In fact, all odd powers of $k$ contain the same overall prefactor.

\subsection{Case II with general $\gamma$: Fragile surface states}

\begin{figure}[H]
\centering
\includegraphics[scale=0.41]{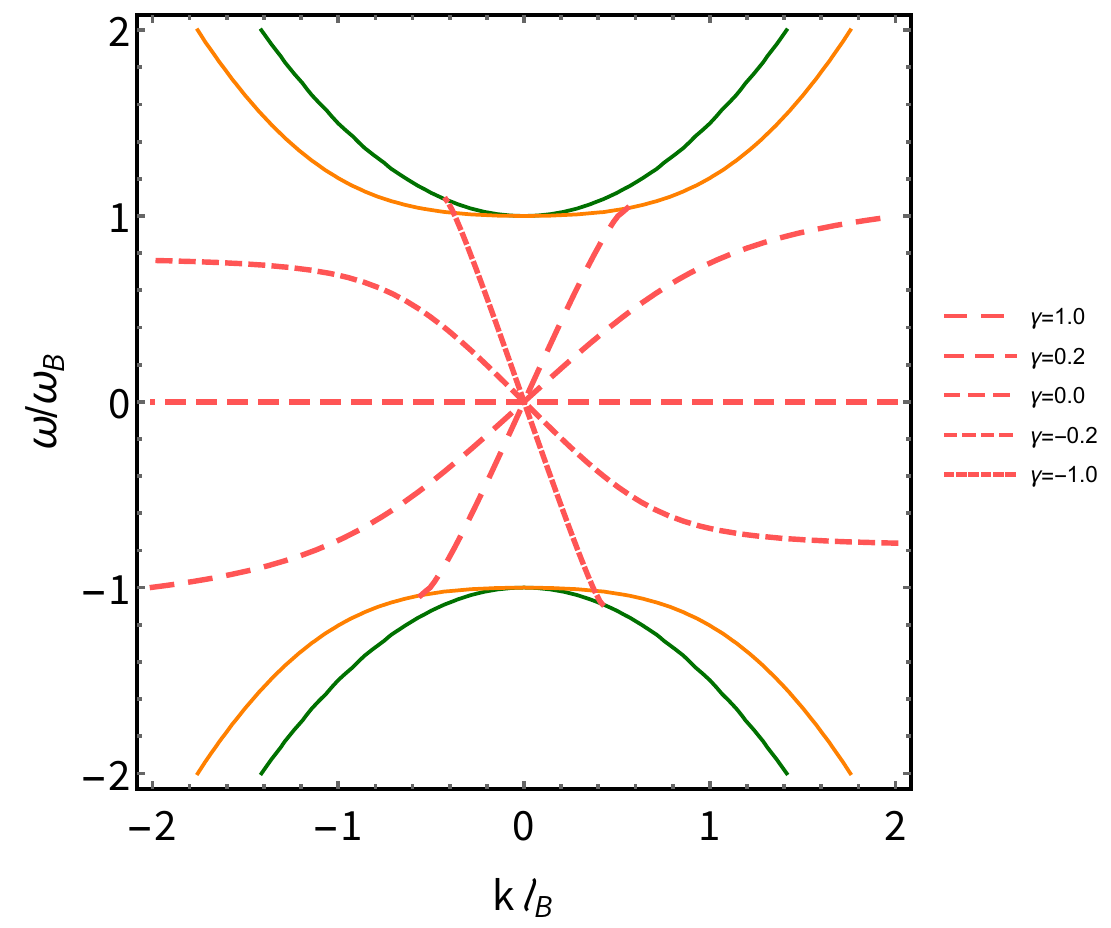}
\caption{Edge modes for various values of $\gamma$, with $V^{11}=V^{22}=0.6\pi\hbar\omega_B\ell_B$, and $V^{12}=V^{21}=0.4\pi\hbar\omega_B\ell_B$.}
\label{fig:Gamma}
\end{figure}

Finally, we analyze the conditions (\ref{eq:gammaV}) and (\ref{eq:gammacont}), which we show does not produce a gapped edge mode. We apply the same technique as above and find the defining relation to be
\begin{align}
\left(\omega-2k\chi^Q\right)(1-\gamma)^2\chi^S - \left(\omega-2k\chi^S\right)(1+\gamma)^2\chi^Q = 0. \label{eq:gammarel}
\end{align}
Here we see that $\gamma=-1$ gives a charge carrying chiral boson mode, and $\gamma=1$ gives a pseudospin carrying chiral boson mode, leading to either (\ref{eq:Qboson}) or (\ref{eq:Sboson}), but not both. For general $\gamma$ this boundary condition interpolates between the two types of edge modes, see Fig. \ref{fig:Gamma} for the full numerical solution. Note that if we multiply both sides of Eq. (\ref{eq:gammarel}) by $1/\gamma^2$ and absorb it into the $(1\pm\gamma)^2$ factor, we see a symmetry $\gamma \rightarrow 1/\gamma$ that preserves the relation. For small $\omega$ and $k$ we find the leading order dispersion
\begin{align}
\omega = \frac{4\gamma c^Q c^S}{(1+\gamma)^2 c^Q + (1-\gamma)^2 c^S} \left(2k - \frac{\ell_B^2}{2} k^3 +\cdots \right) .
\end{align}
Note that for $\gamma=0$ and $\gamma \rightarrow \infty$, the band becomes flat  and this mode detaches from the bulk. Furthermore, as $\gamma$ transitions from the charged chiral boson ($\gamma=-1$) to the pseudospin chiral boson ($\gamma=1$), there must be a detachment and flat band leading to the fragile surface states for $\gamma\neq \pm 1$.

\section{Boundary Conditions from Variational Principle} \label{sec:BC-action}

The set of possible boundary conditions discussed in this work can also be obtained from a variational principle. In the following, we will write an effective edge action that can be added to the original CSGL variables, which, upon varying, yields various hydro boundary conditions discussed in the previous sections. We emphasize that the effective action we derive contains several features of the chiral boson edge theory from the K-matrix formalism. However, this effective action is classical, and to make a more quantitative comparison with the chiral boson theory, we need to quantize this action. Quantization of this effective edge action is beyond the scope of this work and will be considered in a separate publication.

The equations of motion~(\ref{eq:theta}-\ref{eq:alpha}) arise as the saddle points of the action~(\ref{eq:csglaction}). This implies that after varying the action, the ``coefficients" of the field variations on the bulk are set to zero. Keeping the variation of the fields at the boundary unconstrained, the boundary conditions are simply the equation of motion generated by the field variations projected at the boundary. For a fluid restricted to the lower half plane, we observe that the variation of the quadratic part of the action~(\ref{eq:csglaction}), generates the following boundary terms:
\begin{align}
    \delta S_{\text{bulk}}&=-\frac{\hbar}{4\pi }\int dtdx \sum_a\left[\frac{2qB}{m}\delta\theta^a\left(\partial_y\theta^a+\alpha_y^a-\frac{q}{\hbar}A_y\right)\right.\nonumber
    \\
    &+\left.\frac{qB}{2m}\delta n^a\partial_yn^a+\sum_b K^{ab}\left(\delta \alpha_x^a\,\alpha_0^b-\delta\alpha_0^a\,\alpha_x^b\right)\right]_{y=0}. \label{eq:deltaS}
\end{align}
By definition, the bulk variation of the action vanishes on equations of motion. Note that the variation~(\ref{eq:deltaS}) does not generate the boundary conditions discussed in this work. To obtain the hydro boundary conditions discussed in previous sections, we need to add a boundary action to $S_{\text{bulk}}$. Below we consider the edge action associated with the various hydro boundary conditions.

\subsection{No-penetration and no-slip boundary condition}
The no-penetration condition in the original CSGL variables can be expressed as,
\begin{align}
    \left(\partial_y\theta^a+\alpha_y^a-\frac{q}{\hbar}A_y+\frac{1}{2}\sum_b K^{ab}\partial_x n^b\right)\bigg|_{y=0}=0\,. \label{eq:theta-edge}
\end{align}
Note that the $\delta\theta^a$ variation only leads to the first three terms and the last term is missing. Furthermore, for this boundary equation to be consistent with one of the equations of motion (\ref{eq:alpha}), we must also impose that
\begin{align}
    \sum_{b}K^{ab}\left(\partial_t\alpha^b_x-\partial_x\alpha_0^b+\frac{qB}{2m}\partial_x n^b\right)\bigg|_{y=0}=0\,. \label{eq:zeta}
\end{align}
The general solution for this expression can be written as
\begin{align}
    \left(\alpha_0^a-\frac{qB}{2m} n^a\right)\Big|_{y=0}&=\partial_t\zeta^a\,, \label{eq:alphax-edge}
    \\
    \alpha_x^a\Big|_{y=0}&=\partial_x\zeta^a\,, \label{eq:alpha0-edge}
\end{align}
where $\zeta^a$ are general (undetermined) functions.

It is straightforward to show that these equations along with the full form of the no penetration~(\ref{eq:theta-edge}) can be obtained by adding the following boundary action to the system:
\begin{align}
  S_{\text{edge}}&=\frac{\hbar}{4\pi}\int \sum_{a,b}K^{ab}\left[\frac{qB}{m}\left(\partial_x\theta^a+\frac{1}{2}\alpha_x^a-\frac{q}{\hbar}A_x\right)n^b\right.\nonumber
  \\
  &+\left.\partial_t\zeta^a\alpha_x^b-\partial_x\zeta^a\left(\alpha_0^b-\frac{qB}{2m}n^b\right)\right]_{y=0}dtdx\,.
\end{align}

Equation~(\ref{eq:theta-edge}) arises from the variation of $\theta^a$ at the boundary, Eq.~(\ref{eq:zeta}) comes from the variation of the boundary field $\zeta^a$ and Eqs.~(\ref{eq:alphax-edge}-\ref{eq:alpha0-edge}) are obtained from the variation of the gauge field $\alpha_\mu^a$ projected at the boundary.

The second hydrodynamic boundary condition arises from the variation of $n^a$ taken at the boundary, which gives us
\begin{align}
    \left[\sum_b K^{ab}\left(\partial_x\theta^b+\tfrac{1}{2}\alpha_x^b-\tfrac{q}{\hbar}A_x+\tfrac{1}{2}\partial_x\zeta^b\right)-\tfrac{1}{2}\partial_y n^a\right]_{y=0}=0\,.
\end{align}
This expression becomes the no-slip boundary condition for both charge and pseudospin components after using equation~(\ref{eq:alpha0-edge}). The first line in the edge action $S_{\text{edge}}$ is somewhat unsettling since it does not come in the gauge invariant form $\partial_x\theta^a+\alpha^a_x$. Nevertheless, the gauge invariance of this action can be seen explicitly through a field redefinition, that is, after the replacements:
\begin{align}
 \alpha_0^a&= \frac{qB}{2m} n^a+\tilde\alpha_0^a \,
 \\
 \alpha_i^a&=\tilde\alpha_i^a\,
 \\
 \zeta^a&= \tilde\zeta^a+\theta^a\,.
\end{align} 
Therefore, denoting $S_{\text{CSGL}}=S_{\text{bulk}}+S_{\text{edge}}$ the resulting acting assume the familiar form derived in our previous work~\cite{monteiro2022topological}:
\begin{widetext}
 \begin{align}
    S_{\text{CSGL}}&=-\frac{qB}{2\pi}\int \sum_a\left[n^{a}\left(\partial_t\theta^{a}+\tilde\alpha_0^{a}-\frac{q}{\hbar}A_0\right)+\tilde\alpha_0^a+\frac{qB}{2m}n^a+\frac{\hbar}{2m}\left(\partial_i\theta^{a}+\tilde\alpha_i^{a}-\frac{q}{\hbar}A_i\right)^2+\frac{\hbar}{8m}(\partial_i n^{a})^2\right.\nonumber
    \\
     &+\left.\sum_b n^a\left(\frac{\hbar}{4m}K^{ab}\epsilon_{ij}\partial_i\tilde\alpha^b_j+\frac{qB}{2\pi\hbar^2} V^{ab} \, n^b\right)-\frac{\hbar}{2qB}\epsilon_{\mu\nu\kappa}\sum_b K^{ab}(\tilde\alpha_\mu^a+\partial_\mu\theta^a)\partial_\nu\tilde\alpha_\kappa^b\right]d^3x \nonumber
     \\
     &+\frac{\hbar}{4\pi}\int dt dx\sum_{a,b} K^{ab}\left[\tilde\zeta^a\left(\partial_x\tilde\alpha_0^b-\partial_t\tilde\alpha_x^b\right)+\frac{qB}{m}\left(\partial_x\theta^a+\tilde\alpha_x^a-\frac{q}{\hbar}A_x\right)n^b\right]_{y=0}.
\end{align}
\end{widetext}
The no-penetration condition will always be obtained by varying the $\theta^a$ fields, and we will fix this as one of the boundary conditions, as we have done in previous sections. However, there are different possibilities for the second boundary condition. The simplest one is the no-slip condition, which arises naturally from the variation of the fluid density. However, the no-stress, partial slip, and fragile surface state boundary conditions are additional dynamical equations in disguise and require the introduction of auxiliary fields at the boundary, as we will outline below. For the single component Laughlin state, we have shown that the effective edge action includes an auxiliary field with the chiral boson action coupled to the background density at the edge~\cite{monteiro2022topological}. We now develop a generalization of the Laughlin case to the two-component bosonic IQH state. 

\subsection{Effective edge action for no-stress and partial slip boundary condition}
The presence of two components allows for a more general family of energy-conserving boundary conditions. The first case is that of partial slip, where the edge tangent stress of one component induces a tangent velocity or slip at the boundary of the other component. We will now deduce the effective action that generates the partial slip condition. The no-stress condition, where the two edges are completely decoupled, will be obtained by setting the slip length $\lambda^{-1}$ to infinity, that is, $\lambda = 0$. First, we need to express the partial slip condition in its dynamical form, which corresponds to:
\begin{align}
    \sum_b \left[\frac{\hbar}{2m}K^{ab}\left(\partial_t n^b + 2 \partial_x v_x^b\right)+\lambda\,\epsilon^{ab} v_x^b\right]_{y=0}=0\,. \label{eq:lambda-dynamic}
\end{align}
From the last section, we obtained that
\begin{align}
    \delta S_{\text{bulk}}+\delta S_{\text{edge}}=\ldots+\frac{qB}{4\pi}\int dt dx \sum_{a,b}\delta n^a K^{ab}v_x^b\,,
\end{align}
together with the no-penetration condition. Since we do not want to spoil the latter, this additional action must be only a function of $n^a$ and the auxiliary field $\phi^a$. Following the Refs.~\cite{abanov2020hydrodynamics, monteiro2022topological}, we see that such an action must be of the form
\begin{align}
    S_{\text{CB}}&=-\frac{qB}{4\pi}\int dt dx\sum_{a,b}\left[K^{ab}\partial_t\phi^a\left(\partial_x\phi^b+n^b\big|_{y=0}\right)\right.\nonumber
    \\
    &-\left.\frac{m\lambda}{\hbar}\epsilon^{ab}\phi^a\partial_t\phi^b \right].
\end{align}

Combining $S_{\text{CSGL}}+S_{\text{CB}}$, we find that the variation of $n^a$ projected on the boundary gives us 
\begin{align}
    v_x^a\Big|_{y=0}=\partial_t\phi^a\, \label{eq:n-edge}
\end{align}
whereas the equation of motion for $\phi^a$ reads
\begin{align}    \sum_b\left[K^{ab}\left(\partial_tn^b\big|_{y=0}+2\partial_x\partial_t\phi^b\right)+\frac{2m\lambda}{\hbar}\epsilon^{ab}\partial_t\phi^b\right]=0\,.
\end{align}
Upon using Eq.~(\ref{eq:n-edge}), this expression coincides with Eq.~(\ref{eq:lambda-dynamic}).

Note that when $\lambda=0$, we recover the result in 
Ref.~\cite{monteiro2022topological}, which describes the no-stress condition. The limit $\lambda\rightarrow \infty$, can be taken upon redefinition $\phi^a\rightarrow \sqrt{\lambda}\,\phi^a$. This forces the first line in the action $S_{\text{CB}}$ to vanish and we recover the no-slip condition.

\subsection{Effective edge action for fragile surface states}

The one-parameter family of boundary conditions described in equation~(\ref{eq:gammacond}) that leads to the fragile surface states is particularly intriguing. This case can also be derived using a variational principle, offering insights into the origins of these fragile states. With this set of boundary conditions, only one of the equations is dynamical, necessitating just one auxiliary field. By rewriting the second equation in~(\ref{eq:gammacond}) in its dynamical form, we obtain
\begin{align}
    \left[\partial_t(n^1-\gamma n^2)+2\partial_x\left(v_x^1-\gamma v_x^2\right)\right]_{y=0}&=0\,.\nonumber
\end{align}
The above equation is an emergent $U(1)$ symmetry at the edge but the $\gamma$ parameter deforms this definition of the conserved local edge charge $(n^1-\gamma n^2)$ in contrast to the bulk $U(1)$ symmetry which is fixed to be $(n^1\pm n^2)$. Thus the edge $U(1)$ is compatible with the one of the bulk $U(1)$ symmetries only for $\gamma=\pm 1$ for which the edge mode begins and ends at the corresponding bulk band. For values of $\gamma \in (-1,1)$, the edge mode decouples from the edge as shown in Fig.~\ref{fig:Gamma}. The additional edge action that generates the fragile states boundary conditions is given by
\begin{align}
  S_{\text{chiral}}&=\frac{qB}{4\pi}\int dt dx\left[2\gamma\partial_t\phi\partial_x\phi-\partial_t\phi (n^1-\gamma n^2)\big|_{y=0}\right].
\end{align}
Thus, varying the action $S_{\text{CSGL}}+S_{\text{chiral}}$ give us
\begin{align}
    &\delta S_{\text{CSGL}}+\delta S_{\text{chiral}}=\ldots+\frac{qB}{4\pi}\int dt dx \Big[\delta n^1\left(v_x^2-\partial_t\phi\right)  \nonumber
    \\
    &+\left.\delta n^1\left(v_x^1+\gamma\partial_t\phi\right)+\delta\phi\Big(\partial_t(n^1-\gamma n^2)-4\gamma\partial_x\partial_t\phi\Big)\right]_{y=0}.
\end{align}
From the above variations, the boundary conditions ~(\ref{eq:gammacond}) leading to the fragile surface state follow naturally.

\section{Discussion and Outlook} \label{sec:disc}

In this work, we consider a fluid dynamical description of  the integer quantum Hall effect of  bosons modeled by a two-component Chern-Simons-Ginzburg-Landau (CSGL) theory with a mutual Chern-Simons (CS) statistical term with a K-matrix of $K=\begin{pmatrix}
    0 & 1\\ 1 & 0
\end{pmatrix}$ implementing the flux attachment. Contrary to the traditional approach of discarding the bulk bosonic matter and focusing only on the gauge-invariant chiral edge dynamics, we investigate the linearized superfluid hydrodynamics of the bulk bosonic matter subject to energy-conserving boundary conditions. We derive both bulk and edge topological properties within the hydrodynamical framework. We show that the bulk topological invariant is encoded in the fluid polarization algebra in the form of quantized Hall conductivity. We then deduce different kinds of energy-conserving hydro boundary conditions at the hard wall. Since the hydro equations are second-order in derivatives, we need two boundary conditions to determine the full solution.  The first boundary condition is the no-penetration condition $v^a_y\big|_{y=0}=0$ which we do not change across different cases. The second boundary condition has more possibilities and corresponds to four distinct edge dynamics that preserve energy conservation. The first possibility is the no-slip condition $v^a_x\big|_{y=0}=0$ which does not result in chiral edge dynamics. The second case is the no-stress condition, resulting in two counterpropagating chiral edge modes that we identify as chiral bosons disguised as hydro boundary conditions. The third case corresponds to the partial slip condition, where the tangent stress in one of the fluid components generates slip in the second component and vice versa. This case corresponds to the mixing of the two chiral modes, leading to the gapping of the two chiral boson modes. We point out that opening a gap in the quantized chiral boson theory requires nonlinear mixing of the edge modes via a Sine-Gordon term. In terms of the fluid variables, the gapping is achieved in a much more straightforward way.

The last case is the most interesting one and emphasizes the real power of our approach. In this case, we balance the tangent stress between the two layers and balance the tangent velocities (slips) across the two layers with a single parameter. This case results in a single chiral boson mode that, under general conditions, does not begin and end at the bulk bands. This case has been recently reported in Ref.~\cite{altland2024fragility} as fragile surface states manifesting in a non-Wigner-Dyson class of non-interacting topological insulators. It is interesting to note that the fragile surface states manifest in the presence of bulk matter and would be missed by the edge theories deduced from the K-matrix formalism, which does not include any bulk matter. We also obtain a symmetry perspective of the fragility of these edge states in terms of the density associated with the edge 
$U(1)$ symmetry, where we can quantify the precise conditions under which the edge states detach from the bulk bands. We construct effective actions for all these boundary conditions that can be added to the original CSGL theory as the new starting point to understand the bulk and boundary properties of these states. We show the existence of non-anomalous counter-propagating Kelvin modes~\cite{monteiro2023coastal}, which are non-dispersive and seem to be present due to Lorentz invariance within the kelvin mode solution.

In the future, it would be interesting to investigate the general conditions for the presence of these fragile surface states across interacting topological phases that are amenable to the fluid dynamics treatment. We also aim to quantify the microscopic mechanisms that underpin these boundary conditions, which will allow us to construct microscopic lattice models that encode a wider class of boundary phenomena in these topological phases.

\section{Acknowledgments}

We thank Matt Foster for useful comments and discussions.  SG and DR are supported by NSF CAREER Grant No. DMR-1944967. Part of this work was performed
at the Aspen Center for Physics, which is supported by National Science Foundation grant PHY-1607611.

\bibliographystyle{refstyle}
\bibliography{twofluid-Bibliography.bib}

\end{document}